\font\bigtenrm=cmr10 scaled\magstep5
\centerline{\bf{\bigtenrm The Ultimate Future of the Universe,}}
\bigskip \centerline{\bf{\bigtenrm  Black Hole Event
Horizon Topologies, }} 
\bigskip 
\centerline{\bf{\bigtenrm  Holography, and }} 
\bigskip
\centerline{\bf{\bigtenrm The Value of the Cosmological
Constant}}

\medskip

\centerline {by}

\medskip
\centerline {Frank J. Tipler\footnote{$^1$}{e-mail
address: tipler@mailhost.tcs.tulane.edu}}
\centerline {Department of Mathematics and
Department of Physics}
\centerline {Tulane University}
\centerline {New Orleans, Louisiana 70118 USA}
\bigskip

{\bf Abstract}.  Hawking has shown that if black holes were to
exist in a universe that expands forever, black holes would
completely evaporate, violating unitarity.  I argue this means unitarity
requires that the universe exist for only a finite future proper
time.  I develop this argument, showing that unitarity also
requires the boundaries of all future sets to be Cauchy surfaces,
and so no event horizons can exist.   Thus, the null generators of
the surfaces of astrophysical black holes must leave the surface
in both time directions, allowing non-spherical topologies for
black hole surfaces.  Since all information eventually escapes astrophysical
black holes, and since the null surfaces defining astrophysical black holes are
Cauchy surfaces, holography automatically holds.  I further show that unitarity requires the
effective cosmological constant to be zero eventually, since otherwise the universe would
expand forever. 

\bigskip

\centerline{\bf INTRODUCTION}
\bigskip
Hawking showed a quarter century ago that if a black hole were
to exist in a spacetime that exists for infinite proper time, it
would completely evaporate, destroying the information inside
the BH, thereby violating unitarity.  Hawking argued that this
result demonstrated that unitarity was indeed violated, but
since unitarity is absolutely fundamental to quantum mechanics,
I shall explore the implications of assuming that Hawking's
result and unitarity are BOTH correct.  This assumption will be
shown to imply: (1) the universe must be closed, with its future
c-boundary being a single point, which means that there are no
event horizons, and sets of the form $\partial I^+(p)$ for any
event $p$ are Cauchy surfaces, so the information inside a
black hole both gets out in the far future and is also coded
entirely on the surface of an astrophysical black hole:
holography automatically holds; (2) the fact that the generators
of any set $\partial I^+(p)$ eventually leave in the future
direction, in contrast to event horizon null generators, means
that higher genus black hole surface topologies MAY be possible;
(3) the value of the cosmological constant is required to be near
zero; in particular the ``natural value" $\Lambda = 8\pi
c^5/G\hbar$, corresponding the the Planck energy density would
violate unitarity.

\bigskip
\centerline{\bf THE ULTIMATE FUTURE OF THE UNIVERSE }
\bigskip

I shall now show that unitarity strongly constrains the future of
the universe.  Astrophysical black holes exist, but Hawking has
shown that if black holes are allowed to exist for unlimited
proper time, then they will completely evaporate, and unitarity
will be violated.  Thus unitarity requires that the universe must
cease to exist after finite proper time, which implies that the
universe has spatial topology $S^3$.  (All other recollapse
topologies, e.g. $S^2 \times S^1$ [9] and negative $\Lambda$
universes can be eliminated as possibilities by arguments which
will be published elsewhere.)  The Second Law of
Thermodynamics says the amount of entropy in the universe
cannot decrease, but it can be shown ([1], p. 410) that the amount
of entropy already in the CBR will eventually contradict the
Bekenstein Bound [8] near the final singularity unless there are
no event horizons, since in the presence of horizons the
Bekenstein Bound implies the universal entropy $S \leq
constant\times R^2$, where $R$ is the radius of the universe, and
general relativity requires $R \rightarrow 0$ at the final
singularity.  The absence of event horizons by definition means
that the universe's future c-boundary is a single point, call it the
{\it Omega Point}.  MacCallum has shown that an $S^3$ closed
universe with a single point future c-boundary is of measure
zero in initial data space.  Barrow [6] has shown that the
evolution of an $S^3$ closed universe into its final singularity is
chaotic.  Yorke [7] has shown that a chaotic physical system is
likely to evolve into a measure zero state if and only if its
control parameters are intelligently manipulated.  Thus life
($\equiv$ intelligent computers) almost certainly must be
present {\it arbitrarily close} to the final singularity in order
for the known laws of physics to be mutually consistent at all
times.  Misner has shown in effect that event horizon elimination
requires an infinite number of distinct manipulations, so an
infinite amount of information must be processed between now
and the final singularity.  The amount of information stored at
any time diverges to infinity as the Omega Point is approached,
since $S\rightarrow +\infty$ there, implying divergence of the
complexity of the system that must be understood to be
controlled.

\bigskip
\centerline{\bf THE TOPOLOGY OF BLACK HOLE EVENT HORIZONS} 
\bigskip
If event horizons do not exist, then strictly speaking neither
do black holes.  However, astrophysical black holes exist, and I
have constructed [10] a spherically symmetric spacetime
satisfying all the energy conditions which shows how this is
possible:  the spacetime is identical to a dust-filled closed
universe with black holes in the expanding phase, but it has no
event horizons.  So the non-existence of event horizons does not
contradict any observation on astrophysical black holes. 
However, this model does show that we have to take the
theorems (e.g. [3], [4], [5]) proving black hole horizons to be
2-spheres with a grain of salt.  These theorems must in effect
make an assumption about the topology of scri [2], and use the
fact that event horizon generators cannot leave the horizons in
the future direction.  However, if the universe is closed with the
future c-boundary a single point, then it is easy to show that the
boundaries of all future sets which lie in the future of some
Cauchy surface must be Cauchy surfaces.   Thus the null
generators of an astrophysical black hole pseudo-horizon [10]
must leave the black hole surface both in the future and the past,
since I have argued that the universe must have $S^3$ Cauchy
surfaces.  The toroidal horizons of Hughes et al [11] are a slicing
phenomenon in spacetimes with the standard scri [12], and can
exist for only period $\sim M$ because in standard scri
spacetimes, eventually horizon null generators must cease to
enter the horizon, and once on the horizon, can never leave it. 
But neither is necessarily true if only pseudo-horizons exist. 
Thus if unitarity and standard quantum gravity are both true,
higher genus black holes MAY exist.

\bigskip
\centerline{\bf THE VALUE OF THE COSMOLOGICAL CONSTANT} 
\bigskip
Current observations show an accelerating universe
with $\Omega_\Lambda = 2/3$, and if the universe were to
continue to accelerate, black holes would evaporate, violating
unitarity.  Hence, unitarity requires that the acceleration will
eventually return to a de-acceleration, followed by a
recollapse.  Now Gibbons and Hawking [13] have shown that the
vacuum energy in de Sitter space generates thermal Hawking
radiation with a temperature of $T^{deS}_H = (\hbar/2\pi
k_B)\sqrt{\Lambda/3} = 3.9397 \times 10^{-30} h 
\sqrt\Omega_\Lambda$ K, or $T^{deS}_H = 2.25 \times 10^{-30}$
degrees Kelvin with $h = 0.70$ and $\Omega_\Lambda = 2/3$.  So
if the cosmological constant were never to be canceled, any
black hole with a Hawking temperature greater than this would
eventually evaporate and violate unitarity.  Thus only black holes
with a mass greater than $\sim 10^{25}\, M_\odot$ --- more
mass than there is in the visible universe --- could avoid
evaporation.

\medskip
In fact, we can use the same argument to show why the
cosmological constant must be near zero, instead of being its
expected value of $\Lambda = 8\pi G\rho_{vac} = 8\pi
G(c^5/\hbar G^2) = 8\pi c^5/G\hbar$ given by the Planck density. 
If the cosmological constant were this large, in a universe that
expands forever, there would be a finite (though extremely
small) probability that vacuum or other density fluctuations
would give rise to a black hole smaller than the de Sitter horizon
$R^{deS} = c\sqrt{3/\Lambda}$, which would have a higher
Hawking temperature than the de Sitter background temperature,
and hence evaporate, violating unitarity.  So ultimately, the
cosmological constant is near zero because a large cosmological
constant would violate unitarity.

\vfill\eject
\bigskip
\centerline{\bf{References}}
\bigskip
\noindent\hangindent=20pt\hangafter=1
\item{[1]} Tipler F J 1994 {\it The Physics of Immortality} (New
York: Doubleday).

\noindent\hangindent=20pt\hangafter=1
\item{[2]}  Brill, D. R. et al Phys. Rev. D56 (1997), 3600.

\noindent\hangindent=20pt\hangafter=1
\item{[3]} Galloway, G. Comm. Math. Phys. 151 (1993), 53.

\noindent\hangindent=20pt\hangafter=1
\item{[4]} Chrusciel, P. T. and R. M. Wald, Class. Quan. Grav. 11
(1994), L147.

\noindent\hangindent=20pt\hangafter=1
\item{[5]} Jacobson, T. and S. Venkataramani, Class. Quan. Grav. 12
(1995), 1055.

\noindent\hangindent=20pt\hangafter=1
\item{[6]} Barrow, J. D. Phys. Reports 85 (1982).

\noindent\hangindent=20pt\hangafter=1
\item{[7]} Yorke, J. A. et al Phys. Rev. Lett. 68 (1992), 2863.

\noindent\hangindent=20pt\hangafter=1
\item{[8]} Schiffer, M.  and J. D. Bekenstein Phys. Rev. D39 (1989),
1109.

\noindent\hangindent=20pt\hangafter=1
\item{[9]}  Barrow, J. D. and F. J. Tipler, Mon. Not. R. astr. Soc. 216
(1985), 395.

\noindent\hangindent=20pt\hangafter=1
\item{[10]} Tipler, F. J. et al, gr-qc/0003082

\noindent\hangindent=20pt\hangafter=1
\item{[11]}  Hughes, S. A. et al Phys. Rev. D49 (1994), 4004

\noindent\hangindent=20pt\hangafter=1
\item{[12]}  Galloway, G. J. et al gr-qc/9902061

\noindent\hangindent=20pt\hangafter=1
\item{[13]} Gibbons, G. W. and S. W. Hawking, Phys. Rev. 15
(1977), 2738.

\noindent\hangindent=20pt\hangafter=1
\item{[14]} Weinberg, S. Rev. Mod. Phys. 61 (1989), 1.

\vfill\eject
 \bye